# Analytical assessment of the total loss in tubular hollow-core fibers during their fabrication process

Duanny Silva Onório, Flavio A. M. Marques, Alexandre A. C. Cotta, Jefferson E. Tsuchida, Alexandre Bessa dos Santos, Jonas H. Osório

**Hollow-core photonic crystal fibers (HCPCFs) have become a key enabling technology for addressing a broad spectrum of fundamental and applied needs. Indeed, recent advancements achieved by the HCPCF research community have led to significant progress, establishing these fibers as the lowest-loss optical fibers currently available for use in the visible and ultraviolet ranges. However, the fabrication process of HCPCFs demands costly infrastructure, and achieving ultralow-loss fibers remains a complex technical challenge as numerous fabrication attempts are typically required to optimize their performances. Therefore, predicting these fibers' performances before experimental fabrication is highly desirable. In this work, we tackle this task by analytically assessing the total loss in tubular-lattice HCPCFs during their fabrication process. By considering the variation in the microstructure's geometrical parameters during fabrication and the different sources of loss, we estimate expected loss levels and identify the conditions for loss minimization. We understand that our research provides valuable insights into the fabrication process of hollow-core fibers, offering a predictive approach to evaluate the fibers' performance before their experimental realization. By determining optimal conditions considering geometry, fabrication constraints, and loss figures, we believe that our work contributes to the ongoing efforts to further reduce the loss levels in HCPCFs.**

## I. INTRODUCTION

The field of optical fibers has witnessed substantial developments over recent years, driven by the pursuit of enhanced performance across diverse applications ranging from telecommunications to sensing. Among these advances, hollow-core photonic crystal fibers (HCPCFs) have emerged as a breakthrough technology. Indeed, the development of HCPCFs marks a significant milestone in optical fiber technology, positioning these fibers as enablers for advancing various domains within photonics. Thanks to their ability to guide light through an air- or liquid-filled core rather than a solid material, HCPCFs exhibit remarkable potential for a wide range of applications. This capability has driven intense research within the fiber optics community, with efforts focused on optimizing HCPCF designs, refining fabrication techniques, and exploring novel applications. In this framework, HCPCFs are increasingly employed to meet demanding technological needs, for example, facilitating efficient guidance of high-power laser beams [1, 2], enabling optical sources based on nonlinear optics processes occurring in gaseous media [3, 4], besides the realization of fiber sensing configurations [5, 6].

In this context, among the various HCPCF designs investigated in the literature, tubular-lattice HCPCFs – also referred to as antiresonant fibers – have garnered particular attention due to their simplified structure and excellent light guiding performances. Remarkably, tubular HCPCFs stand out for their ability to achieve ultralow loss, especially in the visible and ultraviolet spectral ranges, where they currently represent the lowest-loss optical fibers available [7].

Despite these advances, producing ultralow-loss HCPCFs remains challenging as their fabrication requires precise control over the fiber's structural parameters, hence demanding substantial infrastructure and expertise. Additionally, optimizing the fibers' microstructure during fabrication often entails iterative processes to fine-tune the design parameters, a task that is both resource-intensive and time-consuming. Consequently, predictive tools to assess and optimize HCPCF performance prior to fabrication are highly desirable to streamline development and reduce costs.

In this context, this study addresses the need to have a predictive assessment of HCPCFs loss levels by analytically describing the geometrical parameters of tubular HCPCFs during fabrication to investigate their total loss. Thus, by evaluating the combined impact of the microstructure parameters variations during the fabrication process and applying scaling laws that govern their light-guiding mechanism, we provide a framework for anticipating the fiber loss figures and identifying parameters for optimal transmission conditions regarding fiber loss. Thus, we understand that our efforts offer valuable insights into HCPCF fabrication process, serving as a tool for enhancing the fiber's performances, hence supporting the broader scenario of reducing HCPCF loss levels, besides contributing to amplifying the potential for these fibers to play a transformative role in next-generation photonic applications.

This work was supported by Minas Gerais Research Foundation (FAPEMIG) under the projects APQ-01401-22, RED-00046-23, APQ-01765-22 and APQ-00197-24, and by the National Council for Scientific and Technological Development (CNPq) under grants 305024/2023-0 and 402723/2024-4. Corresponding author: Jonas H. Osório.

Duanny Silva Onório, Flavio A. M. Marques, Alexandre A. C. Cotta, Jefferson E. Tsuchida, and Jonas H. Osório are with the Multiuser Laboratory of Optics and Photonics (LaMOF), Department of Physics, Federal University of Lavras, Lavras, 37.200-900, Brazil (email: jonas.osorio@ufla.br).

Alexandre Bessa dos Santos is with the Faculty of Engineering, Federal University of Juiz de Fora, Juiz de Fora, 36.036-330, Brazil

## II. FIBER GEOMETRY AND LOSS MECHANISMS IN TUBULAR HCPCFs

Fig. 1a displays a representation of the tubular HCPCF microstructure, which is formed by a set of cladding tubes attached to a jacketing tube with internal radius $R_{jac}$. The cladding tubes, with external radius $r_{ext}$ and thickness $t$, define a hollow core with radius $R_{co}$. The distance between adjacent cladding tubes is defined as $\delta$. Such configuration of cladding tubes enables the confinement of light in the fiber hollow core and corresponding light transmission through it. Details on the light guidance mechanism in tubular HCPCFs can be found in [8].

The loss in tubular HCPCFs is typically assessed by using numerical tools (*e.g.*, finite-element method) that allow for estimating the confinement loss (CL) of the modes guided by the fiber taking into account the microstructure design and its corresponding confinement power. Whilst these numerical methods stand for highly powerful approaches for estimating the CL, enabling calculations for different fiber designs, they are typically time-consuming and demand access to commercial software. Considering this, the HCPCF community has made efforts to develop analytical models that could allow for simpler estimation of the CL levels [9-13].

Particularly, in a recent paper, Vincetti and Rosa developed an analytical model for CL estimation in tubular HCPCFs considering the scaling laws in HCPCFs and the fiber microstructure geometrical parameters [13]. By following their methods, we can calculate the CL as a function of the normalized frequency, $F = (2t/\lambda)\sqrt{n^2 - 1}$ ($n$: refractive index of the fiber cladding tubes; $\lambda$: wavelength). For example, we show as a red line in Fig. 1b the calculated CL for a representative fiber with $R_{co} = 20$ μm, $r_{ext} = 10$ μm, $t = 1$ μm, and $\delta = 3$ μm (we refer the reader to the corresponding reference for the complete analytical expressions). Observation of the red curve in Fig. 1b readily reveals the characteristics and behavior of the CL in tubular HCPCF, which encompasses high-loss regions around integer values of $F$ (resonances) and a decreasing loss trend as $F$ increases (smaller wavelengths). For simplicity, here we consider the refractive index of the fiber material (silica) as $n = 1.45$. Also, we mention that we have restricted our analysis to fibers with 8 cladding tubes.

While the CL accounts for the loss that would be expected for a perfect tubular HCPCF, other loss mechanisms must be considered to describe the loss that would be observed in more realistic fibers. In this context, one needs to assess the contributions of other sources of loss, namely the surface scattering loss (SSL, which arises from the surface roughness of the HCPCF core contour generated by thermally-induced surface capillary waves during the fiber drawing [7, 14]), and the microbending loss (MBL, which accounts for the loss arising from the coupling between the core fundamental mode to lossier higher-order modes due to random tilting of the fiber axis during fabrication [15, 16]).

Thus, following analogous efforts to those devoted to avoiding the use of numerical tools for describing the CL, recent endeavors allowed for obtaining expressions for calculating the SSL and MBL analytically [16]. By plugging in the parameters of our representative fiber in the expressions available in [16] (again, we refer the reader to the corresponding reference for the complete expressions of SSL and MBL), we could obtain the curves shown in Fig. 1b for the SSL (dark yellow curve) and MBL (black curve) contributions. Observation of the curves in Fig. 1b reveals the SSL and MBL increasing trends for larger values of $F$ (smaller wavelengths), causing them to greatly exceed the CL contribution for the fiber total loss ($TL = CL + SSL + MBL$, green curve in Fig. 1b) for larger normalized frequencies values. Thus, one analyzes that depending on the targeted frequency range for the fiber operation, different loss mechanisms might dominate. In this context, for lower frequencies (larger wavelengths), one might typically work on the fiber design to achieve lower loss levels (*e.g.*, using fiber designs that provide enhanced confinement power [17, 18]). On the other hand, decreasing the loss in the short-wavelength range (larger frequencies) might consider lowering the SSL and MBL contributions in TL [7].

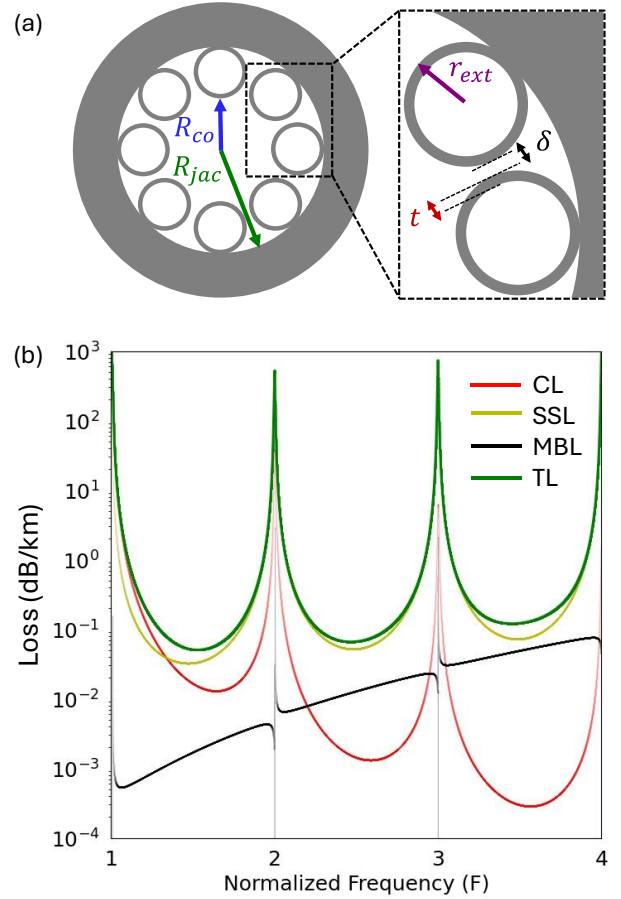

**Fig. 1.** (a) Representation of the tubular HCPCF microstructure with the identification of its geometrical parameters. $R_{jac}$: jacketing tube inner radius; $r_{ext}$: cladding tubes external radius; t: cladding tubes thickness; $R_{co}$: core radius; $\delta$: distance between adjacent cladding tubes. (b) Analytically calculated loss spectrum for a representative tubular HCPCF with the corresponding CL, SSL and MBL contributions to the TL. Here, we considered $R_{co} = 20$ μm, $r_{ext} = 10$ μm, $t = 1$ μm, and $\delta = 3$ μm as representative fiber geometrical parameters.

### III. FIBER GEOMETRICAL PARAMETERS VARIATION DURING FABRICATION

Tubular HCPCFs are typically fabricated using the stack-and-draw method [17]. This process involves arranging capillaries within a jacketing tube according to the desired fiber microstructure design, followed by fiber drawing. During fabrication, the high temperatures required for glass melting combined with surface tension and viscoelastic effects [17] cause the thin cladding tubes in tubular HCPCFs to collapse. To prevent this collapsing trend and ensure the correct tube dimensions within the fiber microstructure, pressurization of the cladding tubes is applied [7, 17]. This technique allows for expanding the tubes during the drawing process, allowing them to reach the desired geometrical dimensions. Fig. 2a illustrates the expansion of the fiber cladding tubes during the fabrication process.

Observation of the diagram in Fig. 3a reveals that the expansion of the cladding tubes during fabrication (accounted here from the variation in $r_{ext}$) induces changes in the other fiber microstructure's geometrical parameters, namely $R_{co}$, $\delta$, and $t$. Using simple geometrical relationships, we can derive Eq. (1) and (2), which describe the connections between the fiber's geometrical parameters, as defined in Fig. 1a ($N$ is the number of cladding tubes) [9]. Furthermore, by considering mass conservation during the fiber fabrication process, we can derive Eq. (3) to express the cladding tube's thickness as a function of its external radius, assuming an initial thickness $t_0$.

$$R_{co} = R_{jac} - 2r_{ext} \quad (1)$$

$$\delta = 2R_{jac} \sin\left(\frac{\pi}{N}\right) - 2r_{ext}\left(1 + \sin\left(\frac{\pi}{N}\right)\right) \quad (2)$$

$$t = r_{ext} - \sqrt{r_{ext}^2 - t_0^2} \quad (3)$$

To study the evolution of the fiber geometrical parameters during fabrication we can, based on Eq. (1), (2), and (3), obtain the graphs shown in Fig. 2b, 2c, and 2d, considering $R_{jac}$ = 40 μm. As expected, Fig. 2b shows that, as the tubes expand, the fiber core radius reduces linearly. In turn, Fig. 2c illustrates the reduction of the spacing between the cladding tubes as they enlarge. In this example, one observes that $\delta$ has been reduced from 23 μm (tubes completely collapsed) to 0.01 μm (tubes at the imminence of touching each other) as $r_{ext}$ changed from 2.6 μm to 11 μm.

Additionally, Fig. 2d shows the evolution of the tube thickness for varying $r_{ext}$. Observation of Fig. 2d reveals the achievable thickness of the cladding tubes according to the initial thickness of it. For example, the green curve in Fig. 2d shows that, for tubes with an initial thickness of 4.6 μm, one can obtain cladding tubes with a minimum thickness of 1 μm for the largest $r_{ext}$ value of 11.1 μm (limit set by the situation in which the tubes touch). In turn, if one considers cladding tubes with an initial thickness of 2.6 μm (black curve in Fig. 2d), the minimum thickness of the cladding tube would be lowered to 0.3 μm.

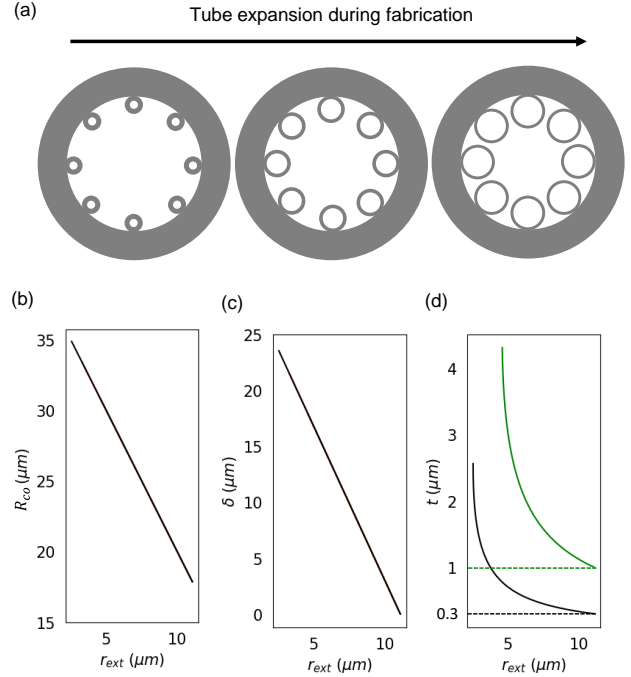

**Fig. 2.** (a) Diagram of the tubular HCPCF cladding tube expansion during fabrication. Representative graphs on the variation of (b) $R_{co}$, (c) $\delta$, and (d) $t$ with varying $r_{ext}$. The plots consider $R_{jac}$ = 40 μm.

The analysis of the plots in Fig. 2 highlights the combined variations in the HCPCF cladding parameters as the microstructure evolves during fabrication. Indeed, these variations in the fiber's geometric dimensions have a complex and nonobvious influence on the different HCPCF loss mechanisms. In the next section, we examine the impact of these cladding modifications on the expected total loss. Additionally, it is worth noting that our analysis, which considers the combined effects of microstructure alterations during fabrication, differs from the conventional approaches reported in the literature. Traditional studies often focus on the variation of individual cladding parameters, which, while providing valuable insights into HCPCF loss mechanisms, do not capture the dynamic evolution of the fiber loss during the fiber drawing process.

### IV. ANALYTICAL ASSESSMENT OF THE FIBER TOTAL LOSS DURING FABRICATION

The evolution of the geometrical parameters during fiber fabrication allows for investigating their corresponding impact on the fiber TL. Indeed, due to the complexity of the loss mechanisms in tubular HCPCFs, accounting for the impact of the geometrical modifications during the fiber draw may be intricate since these parameters are interconnected, so that variations in one of them imply changes on the others, as seen in Eq. (1), (2), and (3). Additionally, the alterations of the geometrical dimensions may have different impacts on the distinct loss mechanisms. For example, when one expands the cladding tubes during fabrication, the reduction of $\delta$ readily enhances the fiber confining power and decreases the CL. On the other hand, expanding the tubes reduces the core radius, which causes the CL to increase. Indeed, one may find similar situations when considering the other sources of loss (besides their corresponding wavelength dependence), making it challenging to anticipate the fiber TL during its fabrication. Moreover, this difficulty is further deepened if one considers that the fiber fabrication procedure is a highly dynamic process, typically involving the need to check the fiber microstructure under a microscope for fine-tuning the fabrication parameters. Therefore, analyzing the impact of the fiber geometrical parameters on the TL as they evolve during fabrication (*i.e.*, as the cladding tubes expand) is highly desirable.

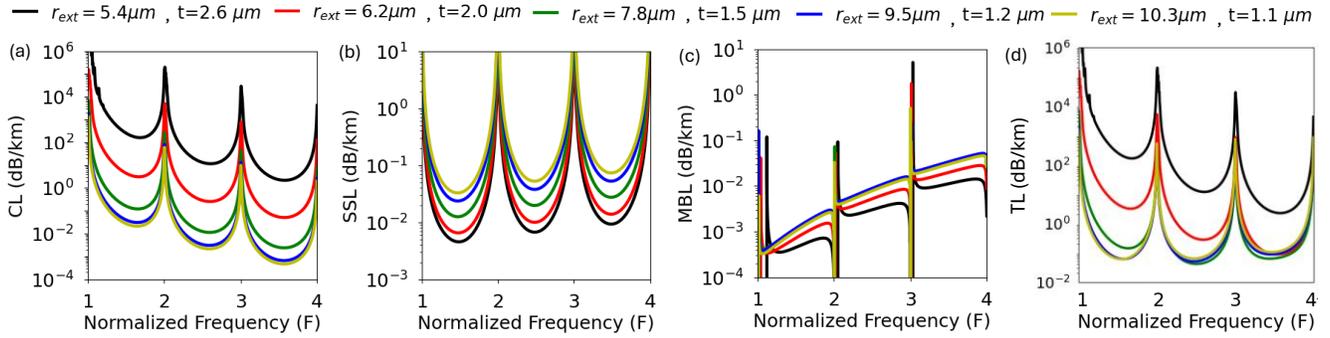

**Fig. 3.** Plots of the (a) CL (b) SSL, (c) MBL and, (d) TL as a function of the normalized frequency considering cladding tubes with varying $r_{ext}$ and $t$ (also reflecting changes in $R_{co}$ and $\delta$), representing their expansion during fiber fabrication. The plots assume $R_{jac} = 40$ μm.

We start by analyzing the impact of the variation of the set of geometrical parameters during fiber fabrication (cladding tube expansion, and corresponding variations in $\delta$, $t$, and $R_{co}$ values) on the loss mechanisms separately. Fig. 3a presents the results corresponding to the fiber CL evolution during fiber fabrication. The expansion of the cladding tubes, whilst it causes $\delta$ to decrease (in this example, from 15.7 μm to 2.2 μm, corresponding to $r_{ext}$ values of 5.4 μm and 10.3 μm, respectively), causes the CL to decrease. Noteworthily, even though the variation of the core radius has a strong impact in CL (the CL minimum values in each transmission band are expected to follow a $R_{co}^{-4}$ trend according to the loss scaling laws in tubular HCPCF [9]), in this example, the effect of reducing $\delta$ dominates that of reducing $R_{co}$ (here, $R_{co}$ varied from 29.2 μm to 19.4 μm).

Additionally, we calculate the evolution of SSL during fabrication as shown in Fig. 3b. Observation of the data in Fig. 4b reveals that the enlargement of the cladding tubes during fabrication contributes to the increase of SSL values. Indeed, this occurs because the SSL is proportional to the overlap between the guided optical mode and the fiber microstructure elements, which is, in turn, proportional to $R_{co}^{-3}$ [9, 16]. Thus, the expansion of the tubes, reducing the core diameter, entails an increase in SSL levels.

In turn, the evolution of MBL during fabrication is shown in Fig. 3c. The results reveal an increase in MBL levels as the tubes expand. Indeed, such an increase can be understood as a consequence of the combined effect of the set of microstructure parameters that are modified during fabrication. Based on MBL scaling laws [16], a reduction on $R_{co}$ (neglecting variations in the other parameters) would imply a decrease in the MBL. However, by evaluating the combined effect of the varying geometry during fabrication, we concluded that, for the parameters studied herein, the decrease in $t$ values due to the tubes' expansion dominated the core radius contribution and caused the MBL to increase.

The previous results allow for obtaining the plot shown in Fig. 3d, which accounts for the TL during the expansion of the cladding tubes. Here, we see that the compromise between the contributions of the different loss mechanisms causes the TL values to stabilize as the tubes are enlarged. Indeed, this stabilization trend is different in each transmission band due to the distinct impact of each loss mechanism. For lower $F$ values, due to the greater influence of the CL, the stabilization occurs for larger values of $r_{ext}$. Otherwise, for larger $F$, thanks to the greater impact of the SSL and MBL, the TL stabilizes for smaller $r_{ext}$, even resulting in a loss augmentation scenario if $r_{ext}$ is further increased (see, for example, that the minimum TL corresponding to $r_{ext} = 9.5$ μm is larger than that for $r_{ext} = 10.3$ μm between $F = 2$ and $F = 3$). Hence, we observe that, depending on the targeted frequency for the fiber operation, one might find an optimum size for the cladding tubes in the fiber microstructure that allows for the TL optimization at a certain $F$.

To illustrate this analysis, we show in Fig. 4 the TL (and its corresponding components) at $F = 1.5, 2.5,$ and $3.5$ for varying $r_{ext}$ (top row of graphs) and corresponding $\delta$ (bottom row of graphs). Specifically, the results corresponding to $F = 1.5, 2.5,$ and $3.5$ are displayed in Fig. 4a, 4b, and 4c, respectively. Observation of the graphs shown in the top row in Fig. 4, readily informs the decreasing trend of CL as $r_{ext}$ is increased (pinkish region). The CL is, therefore, the dominating loss mechanism for smaller cladding tubes due to the corresponding larger $\delta$, which weakens the structure confining power. Alternatively, as the tubes are enlarged during fiber fabrication, the SSL (yellowish region) and MBL (greyish region) contributions arise, causing them to be the dominant loss mechanisms for larger $r_{ext}$.

In this context, the study of the interplay between the loss mechanisms due to the varying microstructure during fabrication allows us to identify the cladding geometrical parameters that minimize the TL in each transmission band. Thus, Fig. 4a shows that, at $F = 1.5$, the TL is minimized at $6.1 \times 10^{-2}$ dB/km for $r_{ext} = 10.1$ μm ($\delta = 2.8$ μm, $R_{co} = 19.6$ μm, $t = 1.1$ μm). In turn, Fig. 4b informs that, at $F = 2.5$, the TL is minimized at $4.0 \times 10^{-2}$ dB/km for $r_{ext} = 8.1$ µm ($\delta = 8.2$ µm, $R_{co} = 23.8$ µm, $t = 1.4$ µm). Accordingly, Fig. 4c reveals that, at $F = 3.5$, the TL is minimized at $5.5 \times 10^{-2}$ dB/km for $r_{ext} = 6.7$ µm ($\delta = 11.4$ µm, $R_{co} = 26.1$ µm, $t = 1.7$ µm). The wavelengths corresponding to the minimized loss at $F = 1.5$, 2.5, and 3.5 are 1555 nm, 1204 nm, and 1040 nm, respectively.

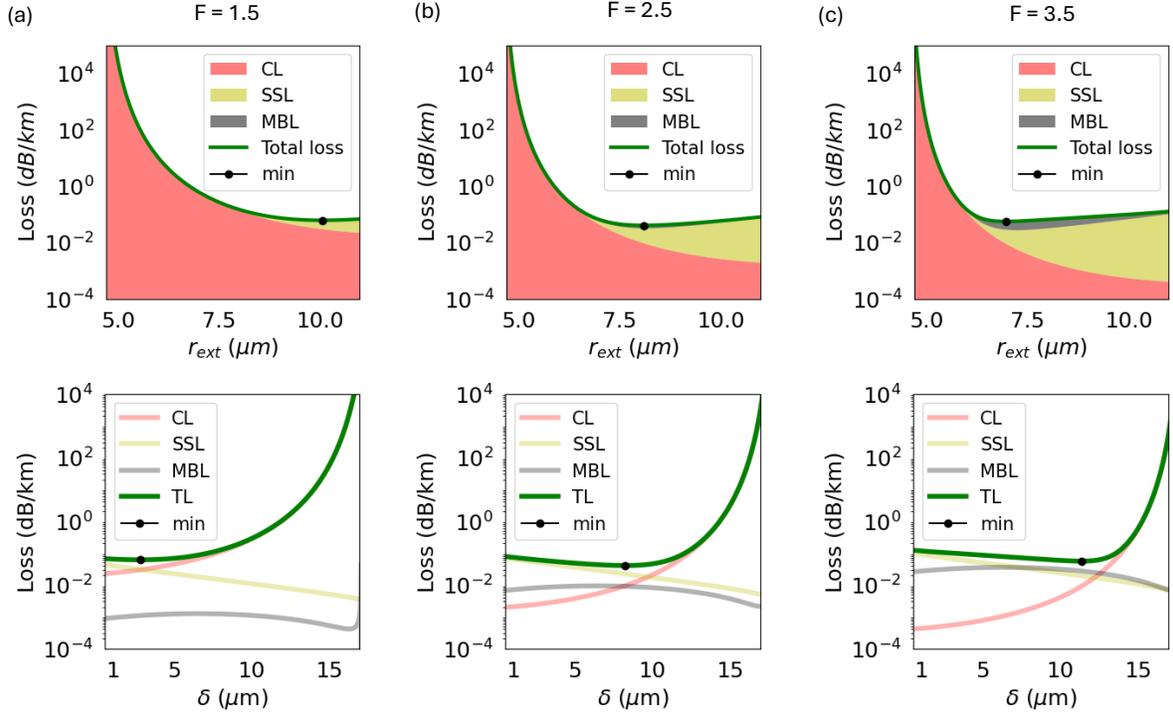

**Fig. 4.** CL, SSL, MBL and TL as a function of $r_{ext}$ (top row) and $\delta$ (bottom row), simulating their variation during fabrication for (a) $F = 1.5$, (b) $F = 2.5$, and (b) $F = 3.5$. In the figures shown, one assumed $R_{jac} = 40$ µm.

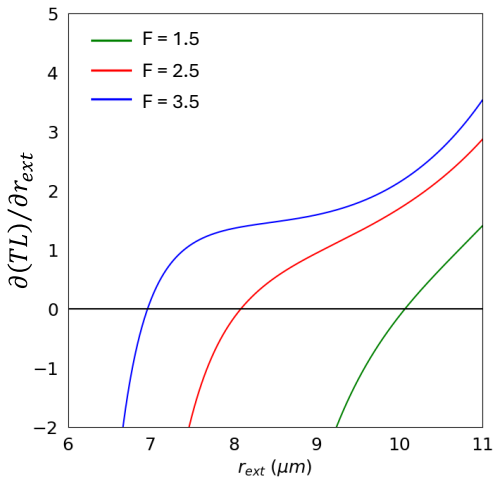

**Fig. 5.** Graph on the partial derivative of the TL with respect to $r_{ext}$, $\partial(TL)/\partial r_{ext}$, as a function of the cladding tubes external radius. This plot allows to readily identify the $r_{ext}$ that minimizes the TL at different normalized frequency values.

The identification of the $r_{ext}$ values that minimize the TL can be further illustrated by the graph shown in Fig. 5, where one plotted the partial derivative of the TL with respect to $r_{ext}$, $\partial(TL)/\partial r_{ext}$, as a function of the cladding tubes radius. The points at which $\partial(TL)/\partial r_{ext}$ is null readily indicate the $r_{ext}$ that entails minimum TL figures, further emphasizing the results shown in Fig. 4. Here, we remark that the above-mentioned TL values may alter if the initial parameters for setting the fiber microstructure or the assumed contributions of the different loss mechanisms (*e.g.*, larger surface roughness elevating SSL levels, different power spectral density describing the MBL) were considered to be different.

Hence, one identifies that, during fiber fabrication, one may target attaining different sizes of the cladding elements to minimize the loss in a considered transmission band. Indeed, the data shown in Fig. 4 and 5 readily informs that when aiming at reducing the TL in spectral regions where SSL and MBL contributions are greater (*i.e.*, at larger $F$), expanding the tubes beyond the point that makes the TL minimum is disadvantageous. Having information on the conditions for having minimum TL values would hence circumvent the need to tackle the fabrication difficulties that arise from having the cladding tubes excessively close to each other (such as avoiding them from touching as the fiber is drawn).

In this scenario, we evaluate that the analysis proposed herein, whilst it can account for the combined variations of the cladding geometrical parameters during fabrication and the corresponding impact on the different loss sources, further allowing for anticipating the minimum loss in each transmission band, contributes to the HCPCF community fiber fabrication efforts by providing the fabricators with valuable approach for identifying to which extent the efforts in having tubes closer to each other would have a significant impact on decreasing the fiber loss levels.

## V. CONCLUSION

The study reported herein presented an analytical framework for evaluating and predicting the total loss in tubular-lattice HCPCFs during their fabrication process. By systematically considering variations in microstructural geometry and key loss mechanisms, we provided insights into the factors influencing fiber performance. We understand that this predictive approach offers a valuable tool for optimizing fabrication parameters, reducing the need for extensive trial-and-error processes, and contributing to the community efforts in achieving ultralow-loss HCPCFs. We believe that the methodologies and results outlined in this work can contribute to future innovations in HCPCF area, supporting upcoming advances in both fundamental research and practical applications using this family of optical fibers.

## REFERENCES

[1] H. C. H. Mulvad, S. A. Mousavi, V. Zuba, L. Xu, H. Sakr, T. D. Bradley, J. R. Hayes, G. T. Jasion, E. N. Fokoua, A. Taranta, S. U. Alam, D. J. Richardson, F. Poletti, "Kilowatt-average-power single-mode laser light transmission over kilometre-scale hollow-core fibre," Nature Photonics 16, 448–453 (2022).

[2] M. A. Cooper, J. Wahlen, S. Yerolatsitis, D. Cruz-Delgado, D. Parra, B. Tanner, P. Ahmadi, O. Jones, M. S. Habib, I. Divliansky, J. E. Antonio-Lopez, A. Schülzgen, R. Amezcua Correa, "2.2 kW single-mode narrow-linewidth laser delivery through a hollow-core fiber," Optica 10, 1253-1259 (2023)

[3] M. Chafer, J. H. Osório, A. Dhaybi, F. Ravetta, F. Amrani, F. Delahaye, B. Debord, C. Cailteau-Fischbach, G. Ancellet, F. Gérôme, F. Benabid, "Near- and middle-ultraviolet reconfigurable Raman source using a record-low UV/visible transmission loss inhibited-coupling hollow-core fiber," Optics & Laser Technology 147, 107678 (2022).

[4] M. I. Suresh, J. Hammer, N. Y. Joly, P. St. J. Russell, F. Tani, "Deep-UV-enhanced supercontinuum generated in a tapered gas-filled photonic crystal fiber," Optics Letters 46, 4526-4529 (2021).

[5] H. Bao, W. Jin, Y. Hong, H. L. Ho, S. Gao, Y. Wang, "Phase-Modulation-Amplifying Hollow-Core Fiber Photothermal Interferometry for Ultrasensitive Gas Sensing," Journal of Lightwave Technology 40, 313-322 (2022).

[6] G. L. Rodrigues, C. M. B. Cordeiro, F. Amrani, F. Gérôme, F. Benabid and J. H. Osório, "High-temperature sensing using a hollow-core fiber with thick cladding tubes," IEEE Sensors Journal 24, 16, 25769-25776 (2024).

[7] J. H. Osório, F. Amrani, F. Delahaye, A. Dhaybi, K. Vasko, F. Melli, F. Giovanardi, D. Vandembroucq, G. Tessier, L. Vincetti, F. Gérôme, F. Benabid, "Hollow-core fibers with reduced surface roughness and ultralow loss in the short-wavelength range," Nature Communications 14, 1146 (2023).

[8] B. Debord, A. Amsanpally, M. Chafer, A. Baz, M. Maurel, J. M. Blondy, E. Hugonnot, F. Scol, L. Vincetti, F. Gérôme, and F. Benabid, "Ultralow transmission loss in inhibited-coupling guiding hollow fibers," Optica 4, 209-217 (2017).

[9] L. Vincetti, "Empirical formulas for calculating loss in hollow core tube lattice fibers," Opt. Express 24, 10313-10325 (2016).

[10] M. Zeisberger, M. A. Schmidt, "Analytic model for the complex effective index of the leaky modes of tube-type anti-resonant hollow core fibers," Scientific Reports 7, 11761 (2017).

[11] Y. Wang, W. Ding, "Confinement loss in hollow-core negative curvature fiber: A multi-layered model," Optics Express 25, 33122-33133 (2017).

[12] M. Bache, M. S. Habib, C. Markos, J. Laegsgaard, "Poor-man's model of hollow-core anti-resonant fibers," Journal of the Optical Society of America B 36, 69-80 (2019)

[13] L. Vincetti, L. Rosa, "A simple analytical model for confinement loss estimation in hollow-core tube lattice fibers," Optics Express 27, 5230-5237 (2019).

[14] P. J. Roberts, F. Couny, H. Sabert, B. J. Mangan, D. P. Williams, L. Farr, M. W. Mason, A. Tomlinson, T. A. Birks, J. C. Knight, and P. St.J. Russell, "Ultimate low loss of hollow-core photonic crystal fibres," Optics Express 13, 236-244 (2005).

[15] E. N. Fokoua, Y. Chen, D. J. Richardson, F. Poletti, "Microbending effects in hollow-core photonic bandgap fibers," 42nd European Conference on Optical Communication, 1-3 (2016).

[16] F. Melli, L. Rosa, L. Vincetti, "Analytical formulas for micro-bending and surface scattering loss estimation in tube lattice fibers," Journal of Lightwave Technology 41, 17, 5714-5721 (2023).

[17] B. Debord, F. Amrani, L. Vincetti, F. Gérôme, F. Benabid, "Hollow-core fiber technology: the rising of gas photonics," Fibers 7, 2, 16 (2019).